\documentstyle[12pt, epsf]{article}
\newcommand{\rhohat}{{\hat{\rho}}}

\newcommand{\bp} {{\bf p}}
\newcommand{\bq} {{\bf q}}

\newcommand{\btab}{\begin{tabbing}}
\newcommand{\etab}{\end{tabbing}}

\newcommand{\beqn}{\begin{equation}}
\newcommand{\eeqn}{\end{equation}}
\newcommand{\barr}[1]{\begin{array}{#1}}
\newcommand{\earr}{\end{array}}
\newcommand{\beqna}{\begin{eqnarray}}
\newcommand{\eeqna}{\end{eqnarray}}
\newcommand{\btablec}{\begin{table} \begin{center}}
\newcommand{\etablec}{\end{center} \end{table}}

\newcommand{\bh}{\beta_{\rhohat}}
\newcommand{\bpi}{\beta_{\pi}}
\newcommand{\brho}{\beta_{\rho}}
\newcommand{\gapproxeq}{\lower.7ex\hbox{$\;\stackrel{\textstyle>}{\sim}\;$}}
\newcommand{\plabel}[1]{\label{#1}}
\newcommand{\pbibitem}[1]{\bibitem{#1}}
\begin{document}
\title{\begin{flushright} \small{hep-ph/9712388}\\
\small{JLAB-THY-97-46} \end{flushright} 
\vspace{0.6cm}  
\Large\bf Photo-- and Electroproduction of $J^{PC}=1^{-+}$ exotics}
\vskip 0.2 in
\author{Andrei Afanasev\thanks{\small \em Current address: North Carolina Central
University, Durham, NC 27707, E-mail: afanas@jlab.org.} \hspace{.1cm}   and   Philip R. Page\thanks{\small \em E-mail:
prp@jlab.org; address from January 15, 1997: T-5, MS-B283, Los Alamos
National Laboratory, P.O. Box 1663, Los Alamos, NM 87545.} \\
{\small \em Theory Group, Thomas Jefferson National Accelerator Facility,}\\ 
{\small \em 12000 Jefferson Avenue, Newport News, VA 23606, USA}}
\date{December 1997}
\maketitle
\begin{abstract}{We estimate the kinematic dependence of the exclusive photo--
and electroproduction of $J^{PC}=1^{-+}$ exotic 
mesons due to $\pi$ exchange.  
We show that the kinematic dependence is
largely independent of the exotic meson form factor, which is
explicitly derived for a $1^{-+}$ isovector hybrid meson in the
flux--tube model of Isgur and Paton. The relevance to experiments
currently planned at Jefferson Lab is indicated.}
\end{abstract}
\bigskip

PACS number(s): 12.39.Mk \hspace{.2cm}12.39.Jh
\hspace{.2cm}12.40.Nn \hspace{.2cm}12.40.Vv \hspace{.2cm} 13.40.Gp \hspace{.2cm}13.60.Le

\section{Introduction}

Evidence for a $J^{PC}=1^{-+}$ isovector state at 1.4 GeV  has
been published most recently in $\pi^{-} p \rightarrow
\eta\pi^{-}p$  by E852 \cite{bnletapi}. Since
the $J^{PC}$ of this state is ``exotic'', i.e. it implies that it is {\it not} a conventional meson,
this has raised significant interest in further experimental
clarification. Specifically, the advent of high luminosity electron
beam facilities like CEBAF at Jefferson Lab have raised the possibility of photo-- or
electroproducing a $J^{PC}=1^{-+}$ state, leading to two conditionally
approved proposals \cite{cebafb1pi,cebafetapi}. 

Experimentally, herculean efforts have been devoted to 
photoproduce $J^{PC}=1^{-+}$ states, but no partial wave analyses have been
reported which would confirm the $J^{PC}$ of the state. 
Condo {\it et al.} claimed an isovector state in $\rho\pi$
with a mass of 1775 MeV and a width of $100-200$ MeV with $J^{PC}$ either
$1^{-+},\; 2^{-+}$ or $3^{++}$ using a 19.3 GeV photon beam \cite{condo}.
Enhancements in $b_1\pi$ have been reported in a similar mass region 
with a photon beam of $25-50$ GeV \cite{omega} and 19.3 GeV
\cite{blackett97}.

In this work we perform the first detailed calculation of the photo--
and electroproduction of $1^{-+}$ states.

\section{Cross--sections}

Since diffractive t--channel exchange is usually taken to be C--parity
even, it follows by conservation of charge conjugation for electromagnetic and
strong interactions that $J^{PC}=1^{-+}$ neutral states cannot be
produced by (virtual) photons $\gamma^{\ast}$ via a diffractive mechanism.  
However, to eliminate the possibility of diffractive exchange
completely, we shall specialize to charge exchange, i.e. to
$\gamma^{\ast}p\rightarrow \rhohat^+ n$, where $\rhohat^+$ is an isovector
state of mass $M_{\rhohat}$ 
with a neutral isopartner with $J^{PC}=1^{-+}$. 

We shall assume in this first orientation that s--channel and u--channel production of states in the mass
range of interest are suppressed, since very heavy $\approx
M_{\rhohat} + M_{p} = 2.5 - 3$ GeV 
excited nucleons need to be produced for this mechanism
to be viable. This leaves us with t--channel meson exchange. 
The lowest OZI allowed mass exchanges allowed by isospin conservation
are $\pi^+,\; \rho^+,\; a_1^+$ and $b_1^+$. Utilizing vector meson
dominance, we note that the $\rho^+$ and $b_1^+$ exchanges require
coupling of $\gamma^{\ast}$ to $\omega$, which is suppressed by 
$(0.30)^2 = 9\%$ relative to the coupling to the $\rho^0$ which occurs
for the other exchanges \cite{page97pho}. Of the remaining exchanges,
$a_1^+$ is likely to be suppressed\footnote{Within Regge phenomenology,
the $a_1$ and $b_1$ are not a leading Regge trajectories.} due to the large mass of the
$a_1^+$ in its propagator. On the other side, $\pi^+$ exchange remains 
possible, and is generally expected to be especially relevant for 
a photon at CEBAF energies. The case is further strengthened
by noting that there is a large $p\pi^+ n$ coupling and that there is
already experimental evidence from E852 for the $\rho^0\pi^+$ coupling of 
a $1^{-+}$ state at 1.6 GeV \cite{bnl97,page97exo}. In contrast, 
$\rho^+$ exchange is expected to be highly suppressed, at least for
hybrid $\protect\rhohat$ in the
flux--tube model, since the relevant coupling $\rhohat^+\rightarrow\omega\rho^+$,
where the photon is regarded as on $\omega$ within VDM, is almost zero
\protect\cite{page95light}. We henceforth
restrict to $\pi^+$ exchange. At CEBAF energies a single particle
rather than a Reggeon picture is appropriate. Nevertheless, we have
verified that a Regge theory motivated $t$ dependence does not
introduce more sizable corrections to our predicted cross--sections
than variations of parameters do.

We write the Lorentz invariant amplitude as
\cite{afanas96}

\beqn  {\cal M} =
e F_{\rhohat\gamma\pi} g_{p\pi n} F_{p\pi n} (t) \frac{i}{M_{\pi}^2-t}
\epsilon_{\mu\nu\alpha\beta}
\epsilon_{\mu}^{\gamma}\epsilon_{\nu}^{\rhohat\ast}
q_{\alpha}^{\gamma} q_{\beta}^{\rhohat} {\bar{u}}_p\gamma_5 u_n
\label{prop}
\eeqn 
where $\epsilon$ denotes the polarization vectors of the incoming $\gamma^{\ast}$
and outgoing $\rhohat$, $q$ is the corresponding 4--momentum, and $u$
is a bispinor for the initial proton and outgoing neutron. The $\pi$
propagator has the form $i/(M_{\pi}^2-t)$, where $t=(q_p-q_n)^2$, and we assume
a conventional monopole form for the cut--off form factor 
$F_{p\pi n} (t) = \Lambda^2 / (\Lambda^2-t)$ with
$\Lambda=1.2$ GeV. We take the nucleon--$\pi$ coupling constant
$g_{p\pi n} = 13.5$ \cite{dumbrajs}. Eq. \ref{prop} is the only Lorentz invariant
structure that can couple the nucleon to a pseudoscalar exchange
(via ${\bar{u}}_p\gamma_5 u_n$), and the pseudoscalar to $\gamma^{\ast}$ and $\rhohat$ 
vector particles. As far as the Lorentz structure is concerned, the
$\pi$ exchange amplitude for virtual Compton scattering
\cite{afanas96}, vector meson 
(e.g. $\rho$) and $\rhohat$ production is identical, since the
amplitude is not dependent on the C--parity or G--parity of the
state. This is the central observation that enables us to link 
$\rhohat$ production with virtual Compton scattering. In fact, 
we suggest that $\rho^{+}$ photo-- and electroproduction should
be able to test the results in this work  directly in the near
future,
since diffractive exchange is not possible. 

Define four (dimensionless) structure functions for the (unpolarized) 
$ep\rightarrow e^{'}\rhohat n$ electroproduction cross--section as \cite{afanas96}

\beqna
\lefteqn{\hspace{-5cm}\frac{d^5 \sigma}{dE^{'}d\Omega_e d\Omega_{\rhohat}} =\frac{\alpha^2}{64\pi^3}\frac{E^{'}}{E}\frac{|\bq^{\rhohat}|}{M_p
W}\frac{1}{Q^2}\frac{1}{1-\epsilon}[\sigma_T+\epsilon 
\sigma_L+\epsilon\cos2\phi\;\sigma_{TT}+\sqrt{2\epsilon
(1+\epsilon)}\cos\phi\;\sigma_{LT}] \nonumber } \\ & &
\epsilon^{-1} = 1+2\;\frac{Q^2+(E-E^{'})^2}{4EE^{'}-Q^2} \label{cros}
\eeqna
where $E(E^{'})$ is the initial (final) electron energy and $\theta_e$
the electron scattering angle in the frame where the proton is at
rest. $M_p$ is the mass of the proton and $\epsilon$ the virtual photon
polarization parameter. $W^2=(q_p+q_{\gamma})^2$ and
$Q^2=-q_{\gamma}^2$. The azimuthal angle $\phi$ and the
$\rhohat$ angle relative to $\gamma^{\ast}$, $\theta_{c.m.}$, are
defined in the centre of mass frame of the target proton and $\gamma^{\ast}$.
>From Eqs. \ref{prop} and \ref{cros} the structure functions are

\beqna 
\lefteqn{\hspace{-2cm}\sigma_T = [(q_0^{\rhohat}|\bq^{\gamma}| - |\bq^{\rhohat}|q_0^{\gamma}\cos\theta_{c.m.})^2 + (|\bq^{\rhohat}|q_0^{\gamma} - q_0^{\rhohat}|\bq^{\gamma}|\cos\theta_{c.m.})^2 + 
  (q_0^{\rhohat})^2|\bq^{\gamma}|^2\sin^2\theta_{c.m.}]\; X \nonumber } \\ & & 
\sigma_L = 2|\bq^{\rhohat}|^2Q^2\sin^2\theta_{c.m.}\; X \nonumber \\ & & 
\sigma_{LT} = 2|\bq^{\rhohat}|\sqrt{Q^2}\;(q_0^{\rhohat}|\bq^{\gamma}| - |\bq^{\rhohat}|q_0^{\gamma}\cos\theta_{c.m.})\;\sin\theta_{c.m.}\; X \nonumber\\ & & 
\sigma_{TT} = -|\bq^{\rhohat}|^2(q_0^{\gamma})^2\sin^2\theta_{c.m.}\;
X \nonumber\\ & & 
X = \frac{-t}{(t-M_{\pi}^2)^2}[F_{\rhohat\gamma\pi} g_{p\pi n} F_{p\pi n}]^2
\plabel{dim} \eeqna
where $q_0$ represents the energies of $\rhohat$ and $\gamma^{\ast}$, and 
$\bq$ the 3--momentum of $\rhohat$ and $\gamma^{\ast}$; all in the
centre of mass frame of the incoming proton and photon.

As we shall see later, the kinematical dependence of cross--sections
will depend only weakly on the $\rhohat$ form factor
$F_{\rhohat\gamma\pi}$. Hence most of the conclusions of this work
depend weakly on the details of the (unknown) form factor, and are
hence independent of the detailed model assumptions made in the next
section. One crucial exception is the absolute magnitude of
cross--sections, which depend strongly on the form factor. 

\section{Flux--tube model form factor for a $1^{-+}$ isovector
hybrid}

A $1^{-+}$ state cannot be a conventional meson due to its quantum
numbers. One possibility is that it is a hybrid
meson. This possibility will be further explored here.
Extensive hybrid meson decay calculations have
been done in the flux--tube model of Isgur and Paton \cite{page97pho,page95light}. The model is
non--relativistic and is formulated in the rest frame of the
hybrid. Since the hybrid form factor is Lorentz invariant it
can be evaluated in any frame, particularly the hybrid rest frame. The
Lorentz invariant relativistic amplitude, evaluated in the hybrid rest
frame for a hybrid of polarization 1, is \cite{dumbrajs}

\beqna \plabel{rel1}
\lefteqn{{\cal M}_{R} =  e F_{\rhohat\gamma\pi}
\epsilon_{\mu\nu\alpha\beta}
\epsilon_{\mu}^{\gamma}\epsilon_{\nu}^{\rhohat\ast}
q_{\alpha}^{\gamma} q_{\beta}^{\rhohat} 
= -i e F_{\rhohat\gamma\pi} M_{\rhohat} |{\bf p}_{\gamma}| \nonumber }\\ & & 
{\cal M}_{R} =
\sqrt{2E_{\pi}\; 2E_{\rho}\; 2M_{\rhohat}}\; {\cal M}_{NR}\plabel{rel2}
\eeqna
where we wrote the relativistic amplitude in terms of the
non--relativistic amplitude which we shall compute. The
meson wave functions are normalized differently in a non--relativistic
model than in a relativistic case as shown in Eq. \ref{rel2}. Here
$E_{\pi}$ and $E_{\rho}$ (from $\gamma^{\ast}$ via VDM) are the
on--shell energies of the $\pi$ and $\rho$, each with momentum
$|\bp_{\gamma}|$ in the hybrid rest frame. 

The evaluation of the non--relativistic amplitude proceeds as follows.
It is taken to be the product of the VDM coupling of $\gamma^{\ast}$ to the $\rho$,
the propagator of the $\rho$ and the  flux--tube model amplitude for the decay
of a $1^{-+}$ hybrid to $\rho\pi$

\beqn\plabel{vec}
{\cal M}_{NR} =
\frac{e}{2\gamma_{\rho}}\frac{M_{\rho}^2}{M_{\rho}^2+Q^2}\;\;\mbox{Flux--tube
model amplitude}
\eeqn
where $\gamma_{\rho} = 2.52$ \cite{page97pho}. The
flux--tube model amplitude is evaluated as enunciated in by Close and
Page \cite{page95light}, i.e. we assume S.H.O. wave functions for
the $\rho$ and $\pi$, with the hybrid wave function and the flux--tube
overlap as in ref. \protect\cite{page95light}, except that the small
quark--antiquark seperation $r$ behaviour of the hybrid wave function
is $\sim r$.

Utilizing Eq. \ref{rel1} to express the form factor in terms of the
relativistic amplitude, and  to write this in terms of the
non--relativistic amplitude; and using Eq. \ref{vec}, we obtain

\beqna
\lefteqn{\hspace{-3cm} F_{\rhohat\gamma\pi} =
\frac{32\;\pi^{\frac{3}{4}}}{\gamma_{\rho}}
\left(\frac{\sqrt{{\bf p}_{\gamma}^2+M_{\pi}^2}\sqrt{{\bf p}_{\gamma}^2+M_{\rho}^2}}{M_{\rhohat}}\right)^{\frac{1}{2}}\frac{M_{\rho}^2}{M_{\rho}^2+Q^2}
\; \frac{0.62\; \gamma_0}{(1+\frac{0.2}{\bpi^2+\brho^2})^2}
\nonumber }\\ & & \times 
\frac{ (\bpi\brho)^{\frac{3}{2}}\;\bh^{\frac{5}{2}}\; (\bpi^2-\brho^2) }{ (\bpi^2+\brho^2)^{\frac{5}{2}}\;\xi^{\frac{5}{2}} }
\exp(-\frac{{\bf p}_{\gamma}^2}{4\xi}) \nonumber \\ & & 
\xi = 2\bh^2+ \frac{1}{2}
(\bpi^2+\brho^2)-\frac{1}{2}\frac{(\bpi^2-\brho^2)^2}{\bpi^2+\brho^2} \plabel{flux}
\eeqna
up to a sign.
Notice that the pair creation constant $\gamma_0$ of the $^3P_0$ model
enters explicitly in Eq. \ref{flux}. This is because the flux--tube
model, within the assumptions made for the wave functions, gives a
prediction for the couplings of a hybrid in terms of couplings for
mesons in the $^3P_0$ model \cite{page95light}. We use $\gamma_0=0.53$ which reproduces
conventional meson decay phenomenology \cite{geiger94}.
In Eq. \ref{flux}, $\beta$ refers to the inverse radius of the state,
the parameter that enters in the wave function. Due to the 
$\bpi^2-\brho^2$ term, we note that if $\bpi=\brho$ the form factor
vanishes, which explicitly exforces the selection rule that hybrid
coupling to two S--wave mesons is suppressed \cite{page95light}.

\begin{figure}[t]
\plabel{fig1}
\let\picnaturalsize=N
\def\picsize{3in}
\def\picfilenamea{cap1.ps}
\ifx\nopictures Y\else{\ifx\epsfloaded Y\else\input epsf \fi
\let\epsfloaded=Y
\centerline{
\ifx\picnaturalsize N\epsfxsize \picsize\fi \epsfbox{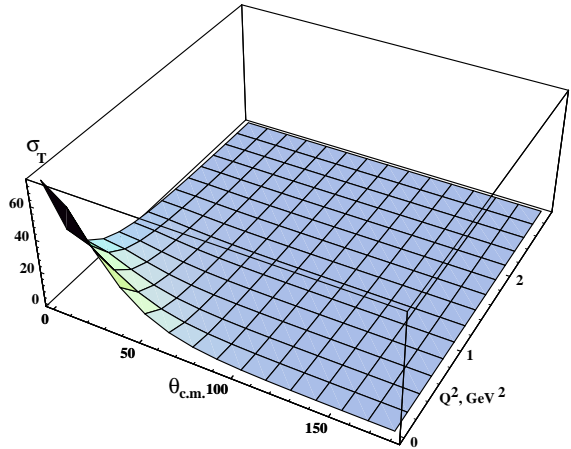}}}
\fi

\caption{Structure function $\sigma_T$ at $W=3$ GeV with standard
parameters. $Q^2$ is varied within its kinematically allowed range for
$E=6$ GeV.}
\end{figure}

\section{Electroproduction results}

We utilize the ``standard parameters''
$M_{\rhohat} = 1.8$ GeV, $\bh= 0.27$ GeV \protect\cite{page95light}, $\brho=0.31$
GeV and $\bpi=0.54$ GeV \protect\cite{swanson92}.

All the kinematical variables that the structure functions depend on,
introduced in Eqs. \ref{cros}, 
\ref{dim} and \ref{flux}, can be expressed as functions of the Lorentz invariant
variables $Q^2$ and $W$, and $\theta_{c.m.}$ (see Appendix).

\begin{figure}[t]
\plabel{fig2} 
\let\picnaturalsize=N
\def\picsize{3in}
\def\picfilenamea{cap2a.ps}
\ifx\nopictures Y\else{\ifx\epsfloaded Y\else\input epsf \fi
\let\epsfloaded=Y
\centerline{
\ifx\picnaturalsize N\epsfxsize \picsize\fi \epsfbox{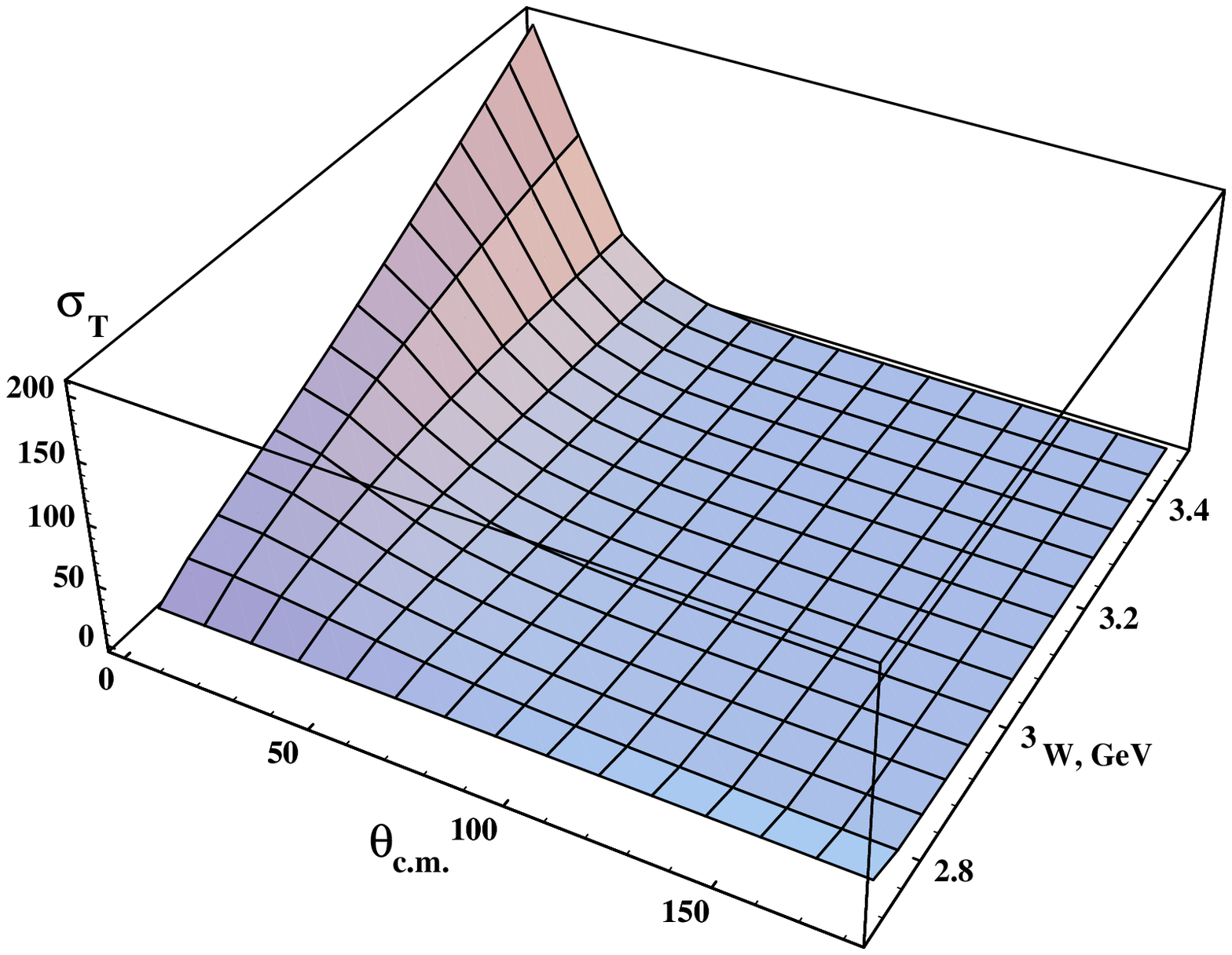}}}
\fi
\let\picnaturalsize=N
\def\picsize{3in}
\def\picfilenamea{cap2b.ps}
\ifx\nopictures Y\else{\ifx\epsfloaded Y\else\input epsf \fi
\let\epsfloaded=Y
\centerline{
\ifx\picnaturalsize N\epsfxsize \picsize\fi \epsfbox{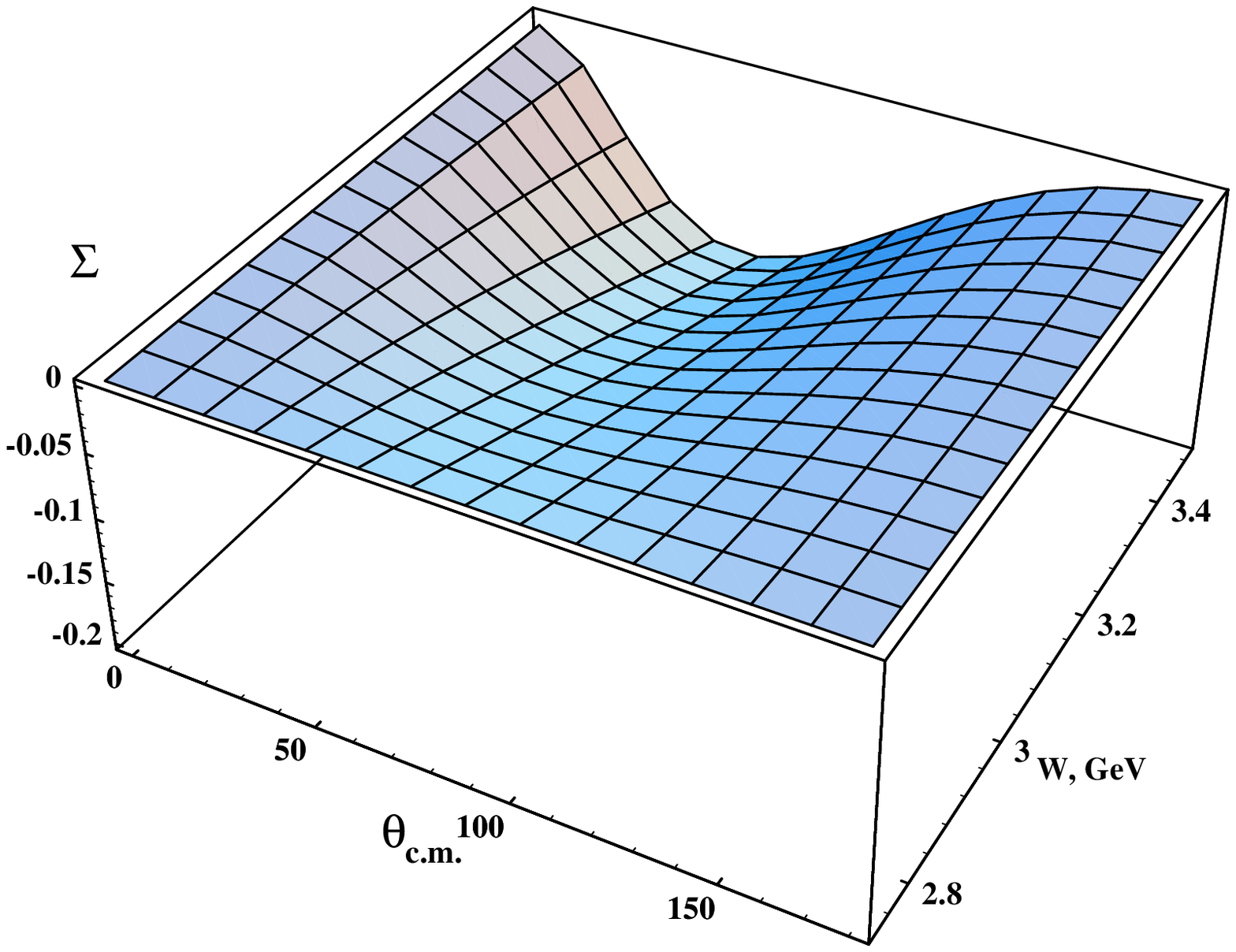}}}
\fi

\caption{Structure function $\sigma_T$ for
$Q^2=0$ and the $\Sigma$  asymmetry of photoproduction with standard parameters.
$W$ is varied within its kinematically allowed range  for
$E=6$ GeV.
}
\end{figure}

The structure function $\sigma_T$  is plotted in
Figure 1 for $W=3$ GeV. $\sigma_T$ is the most dominant structure
function: it peaks strongly at small $Q^2$ and
$\theta_{c.m.}$. Physically, $Q^2 = 0$ corresponds to the incoming and
outgoing electrons moving in the same direction. $\theta_{c.m.}=0$
corresponds to the photon and the $\rhohat$ moving in the same
direction. Hence $\sigma_T$ peaks where the $\rhohat$ goes in
the same direction as the incoming electron, i.e. towards the beam pipe. This becomes especially critical when there is a sizable ``hole'' in
the detector, which is the case for the CLAS spectrometer at
CEBAF. The other three structure
functions are small
when compared to $\sigma_T$, with a suppression factor of about $10^{-3}$
for $\sigma_L$ and
$10^{-2}$ for $\sigma_{LT}$ and $\sigma_{TT}$. These three structure
functions also peak at small $Q^2$ and $\theta_{c.m.}$. 
Experiments should be optimized to enable detection at small $Q^2$ and
$\theta_{c.m.}$. 
According to the Appendix (Eq. \ref{tdep}), $\theta_{c.m.}=0$
corresponds to the minimal value of $|t|$, so that peaking of
cross--sections at small $|t|$ would be a strong experimental test for
the $\pi^+$ exchange explored here, especially since other exchanges
are expected to be more substantial at larger $|t|$.

As we pointed out, the structure function due to
longitudinal photons $\sigma_L$ is tiny. Correspondingly,
$\sigma_{LT}$ which is due to interference between longitudinal and
transverse photons is smaller than $\sigma_T$. The reason for this is
that longitudinal photons give no contribution to the process in a
typical case: when $\rhohat$ is at rest the amplitude $\epsilon_{\mu\nu\alpha\beta}
\epsilon_{\mu}^{\gamma}\epsilon_{\nu}^{\rhohat\ast}
q_{\alpha}^{\gamma} q_{\beta}^{\rhohat}$ in
Eq. \ref{prop} vanishes. The suppression of contributions from
longitudinal photons need not be true for exchanges other than $\pi^+$
exchange.


\begin{figure}[t]
\plabel{fig3}
\let\picnaturalsize=N
\def\picsize{3in}
\def\picfilenamea{cap3.ps}
\ifx\nopictures Y\else{\ifx\epsfloaded Y\else\input epsf \fi
\let\epsfloaded=Y
\centerline{
\ifx\picnaturalsize N\epsfxsize \picsize\fi \epsfbox{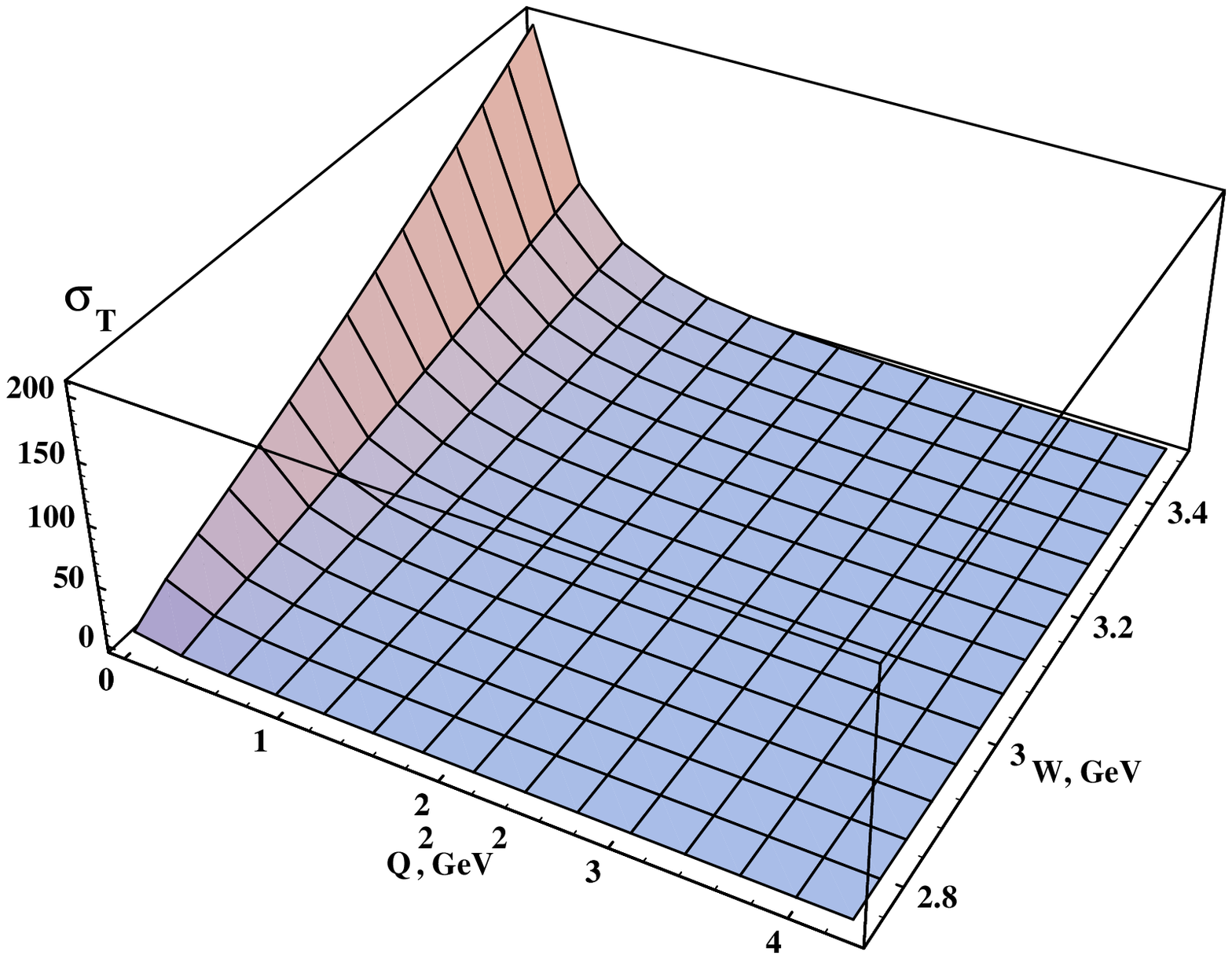}}}
\fi
\caption{Structure function $\sigma_T$ for
$\theta_{c.m.}=0$ with standard parameters. If a curve is drawn from
the top to the bottom corner described by $Q^2 = Q^2_{max}$ (see Eq. \protect\ref{max}), then the region to the
left of the curve corresponds to the physically accessible region of
parameter space for $E=6$ GeV.
}
\end{figure}

In Figure 2 we show the non--zero structure functions for
$Q^2=0$ corresponding to real (transversely polarized) photons.
Again $\sigma_T$ is dominant. Both $\sigma_T$ and $\sigma_{TT}$ peak
at large $W$ as would be expected because large $W$ corresponds to an
increase of phase space for the production of the $\rhohat$. 
We also plot the photoproduction asymmetry parameter $\Sigma = \sigma_{TT} /
\sigma_T$, which can be accessed by using linearly polarized photons.
Note that $\Sigma=0$ at the reaction threshold.

\begin{figure}[t]
\plabel{fig4}
\let\picnaturalsize=N
\def\picsize{3in}
\def\picfilenamea{cap4a.ps}
\ifx\nopictures Y\else{\ifx\epsfloaded Y\else\input epsf \fi
\let\epsfloaded=Y
\centerline{
\ifx\picnaturalsize N\epsfxsize \picsize\fi \epsfbox{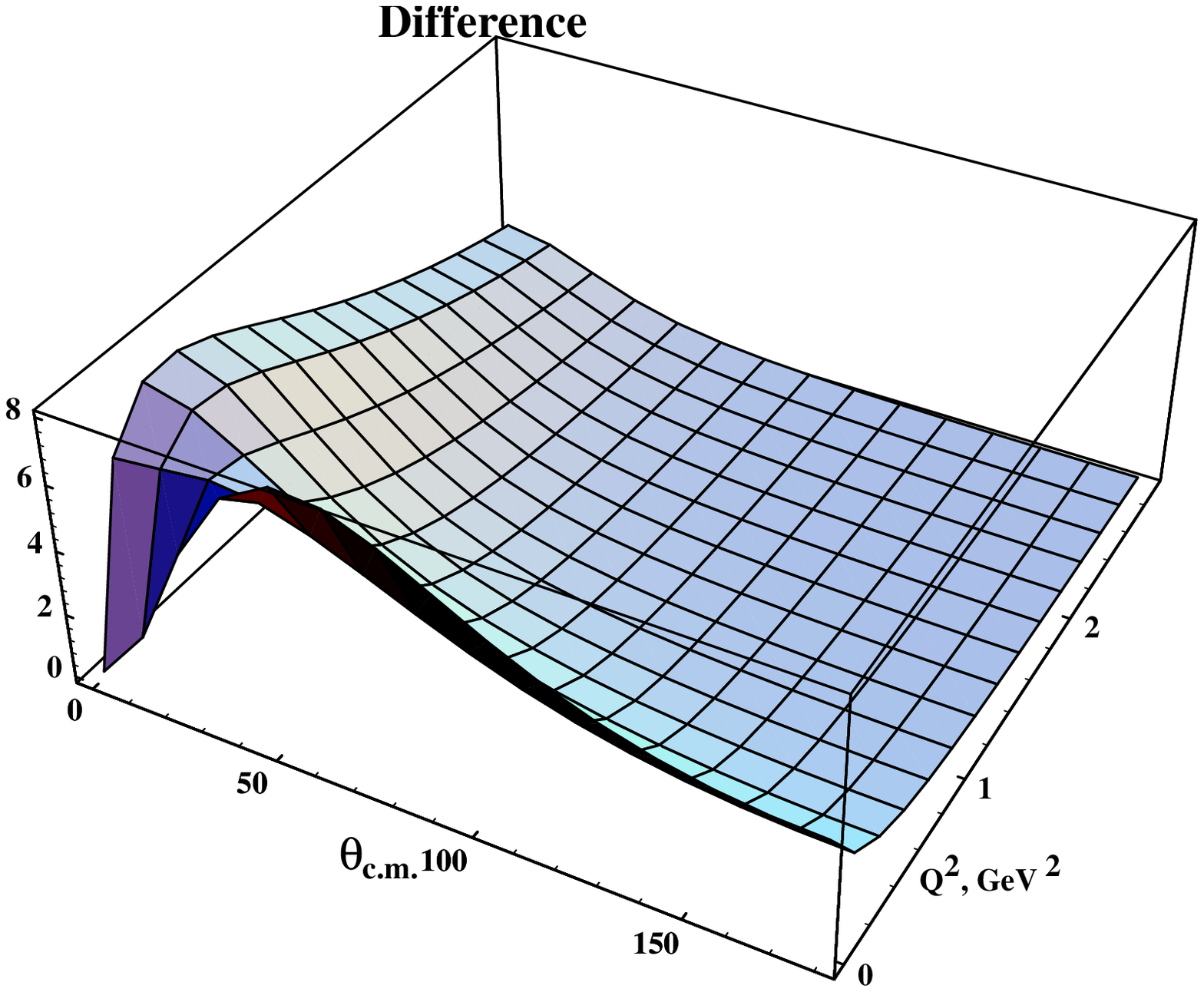}}}
\fi
\let\picnaturalsize=N
\def\picsize{3in}
\def\picfilenamea{cap4b.ps}
\ifx\nopictures Y\else{\ifx\epsfloaded Y\else\input epsf \fi
\let\epsfloaded=Y
\centerline{
\ifx\picnaturalsize N\epsfxsize \picsize\fi \epsfbox{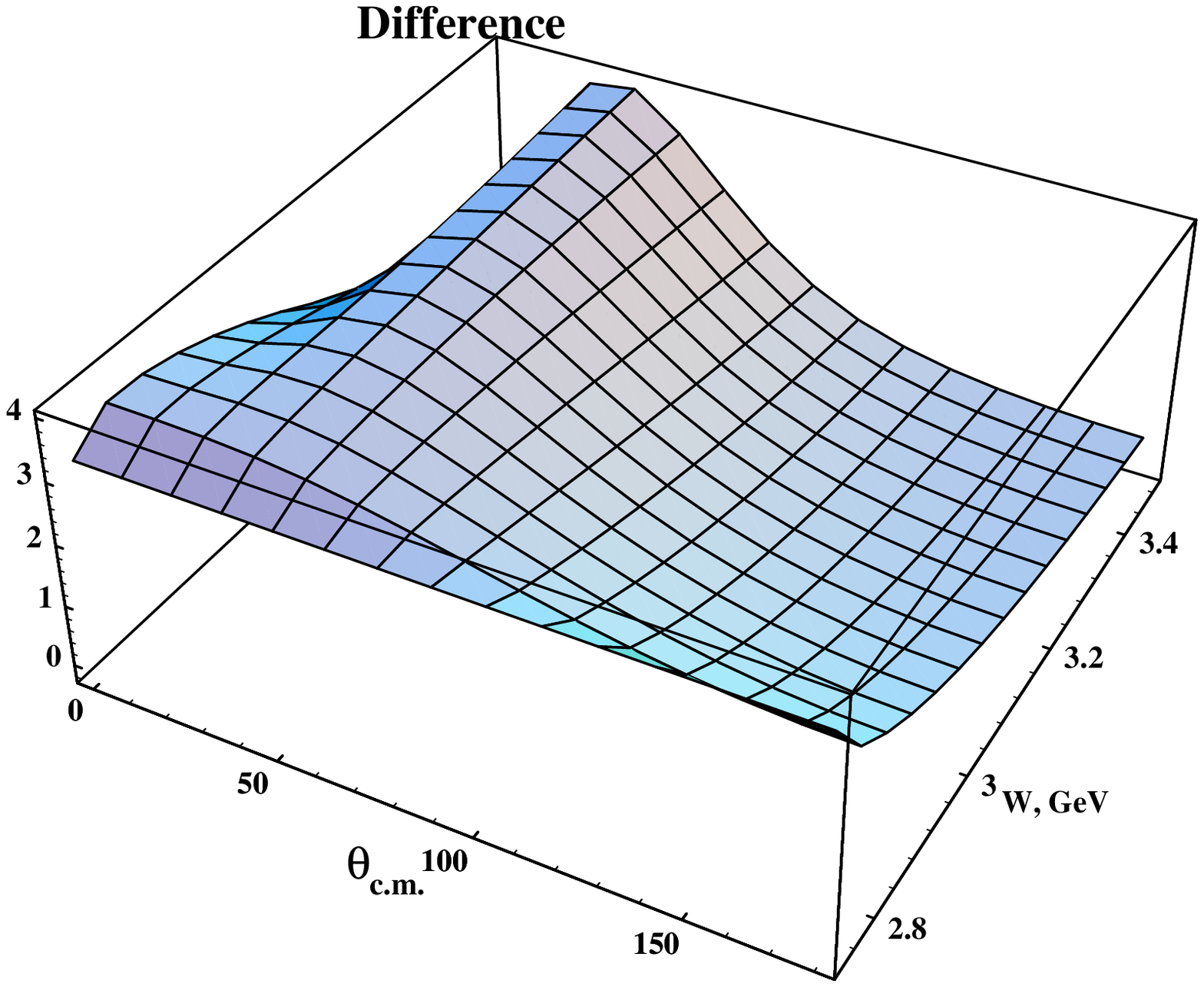}}}
\fi
\caption{The difference (in percent) between the  structure
function $\sigma_T$ with the test
form factor (denoted $\sigma_T^{{{ test}}}$) and
$\sigma_T$ with the flux--tube model form factor  for
standard parameters
as a function of $Q^2, W$ and $\theta_{c.m.}$, varied within their
kinematically allowed ranges for $E=6$ GeV. We normalize
the test form factor to agree with the flux--tube model form factor at points where $\sigma_T$ is maximal,
denoted $\sigma_{{{ max}}}$. For
the first graph $W=3$ GeV, and for the second $Q^2=0$. The ``difference''
is defined as $(\sigma_T^{{{ test}}}
\;\sigma_{{{ max}}} /
\sigma_{{{ max}}}^{{{ test}}}\; - \sigma_T)/\sigma_{{{ max}}}$.
}
\end{figure}

Figure 3 shows the non--zero structure function for
$\theta_{c.m.}=0$, where $\sigma_T$ attains its maximum and the negative value $t\neq 0$ is nearest
to $0$.   

We define a typical test form factor based on $\rho$ dominance as

\beqn\plabel{vir}
F_{\rhohat\gamma\pi} \propto \frac{1}{M_{\rho}^2 + Q^2}
\eeqn
We have evaluated the structure functions for the test form
factor in Eq. \ref{vir}. Remarkably, for all values of $Q^2, W$ and
$\theta_{c.m.}$ the form of the structure functions are very similar,
even though the form factors in Eqns. \ref{flux} and \ref{vir} have
different functional dependence on different parameters.
This is demonstrated for the
dominant structure function $\sigma_T$ in Figure 4, where we
see that the difference is a few percent. Thus the
$Q^2, W$ and $\theta_{c.m.}$ dependence of the cross--section in
Eq. \ref{cros} is very weakly dependent on models for
$F_{\rhohat\gamma\pi}$, so that the kinematic dependence of total
cross--sections, and hence many conclusions of this work, are
independent of the details of specific models. This happens because the
Lorentz structure of one $\pi$ exchange (Eq. \ref{prop}), and not the form factor, governs
kinematical dependence.

We shall now evaluate the total cross--section by integrating over all
kinematical variables in their allowed ranges, except for the
following. The electron scattering angle $\theta_e$ is assumed to be
larger than $\theta_e^{min}$, and $E^{'}$ is assumed to be larger than
$0.1$ GeV. From a theoretical viewpoint, these conditions ensure that
we do not reach $\theta_e=0$ and $E^{'}=0$ where the cross--section 
in Eq. \ref{cros} diverges. Experimentally, the
outgoing electron is usually detected for $\theta_e>\theta_e^{min}$. There are
also experimental limits on detection of small outgoing electron energies.

For the total cross--section, the results are shown in Table \ref{table1}.
The decrease of cross--section for increased $\rhohat$ mass is due to the
decrease in available phase space. The decrease of cross--section with
increasing electron energy is due to the ``hole'' in the forward
direction through which an ever increasing number of electrons
pass. The qualitative dependence of the cross--section on 
$E$ is also found for the test form factor, and is hence mostly model independent. One of the implications of Table \ref{table1} is
that for the CLAS detector at CEBAF, an electron beam towards the
lower end of the range (e.g. 5.5 GeV) appears to be
preferable. Another implication is that at DESY HERA with a 27.52 GeV
proton beam and 820 GeV electron beam, corresponding to $E=48.1$ TeV,
$1^{-+}$ $\rhohat$ production should be negligible.

\begin{table}[t]
\begin{center}
\caption{Total electroproduction cross--section in $pb$ for $\theta_e^{min}=12^o$ and $E^{'}$ larger than
$0.1$ GeV, relevant to the CLAS detector at CEBAF. We utilize the
standard parameters. } 
\label{table1}
\begin{tabular}{|l||r|r|r|}
\hline 
Electron Energy & \multicolumn{3}{|c|}{$\rhohat$ Mass } \\
(GeV)           & 1.4 GeV & 1.8 GeV & 2.2 GeV \\
\hline \hline 
5.5   &  62          & 29        & 3.7\\
6     &  50          & 28        & 6.5\\
6.5   &  41          & 25        & 7.9\\
8     &  21          & 16        & 7.8\\
20    &   0.6        &  0.5      & 0.4\\
\hline 
\end{tabular}
\end{center}
\end{table}

We have also computed the total cross--section for various values of $\theta_e^{min}$ and
obtain

\begin{centering}
\begin{tabbing}
\sl $\theta_e^{min}$ \hspace{1cm} \= Total cross--section ($pb$)\\
$12^o$ \>   28       \\
$ 5^o$ \>  110       \\
$ 1^o$ \> 360         
\end{tabbing}  
\end{centering}
so that the cross--section increases substantially as the ``hole'' in
the detector becomes smaller. This implies that improved statistics
for $\rhohat$ should result from the ability to put detectors as near
as possible to the beam pipe in the forward direction.

It is of interest to check the total cross--section as a function of
the wave function parameters of the participating conventional mesons for $\theta_e^{min}=12^o$. 

\begin{tabbing}
\hspace{8cm} \= Total cross--section ($pb$)\\
Standard parameters \>  28 \\
$\brho=0.45$ GeV and $\bpi=0.75$ GeV \protect\cite{kokoski87} \>  15
\end{tabbing}  
We note that the cross--section changed by a factor of two if the two
conventional meson wave function parameters are changed to reasonable
values.    Also, we chose a value of $\gamma_0$ towards the upper end
of the range in the literature \cite{geiger94}. In calculations of
excited mesons, values of $\gamma_0^2$ that are 50\% lower have been used. Hence, 
within this model, revisions in the $\beta$'s and $\gamma_0$ can make
the cross--sections $\sim 30\%$ of the values quoted for electroproduction
cross--sections in this section and Table 1. 
Hence absolute cross--sections should be regarded with
more caution than kinematic dependence. 

To summarize this section, we stress that the $t$-channel $\pi$ exchange mechanism of $1^{-+}$ electroproduction
leads to dominance of transverse photoabsorption. Therefore a Rosenbluth--type separation of different structure 
functions contributing to the cross section would be necessary in order  to
understand the $\rhohat$ electroproduction  mechanism.

\section{Photoproduction results}

{\normalsize The photoproduction cross--section }($Q^2=0$){\normalsize \ is }

{\normalsize 
\begin{equation}
\frac{d\sigma _\gamma }{d\Omega _{c.m.}}=\frac \alpha {16\pi }\frac{|\bq^{\rhohat}|}{M_p(W^2-M_p^2)}(\sigma _T+\epsilon \cos 2\phi  \sigma _{TT}),
\end{equation}
where }$\phi $ {\normalsize is the angle defined by the planes of photon
linear polarization and $\rhohat$ production; and the parameter $\epsilon$ defines the degree of photon linear polarization.
The total photoproduction
cross--section may be obtained by integrating the preceding formula over }$%
\Omega _{c.m.}${\normalsize ,} {\normalsize 
\begin{equation}
\sigma _\gamma=\frac \alpha 8\frac{|\bq^{\rhohat}|}{M_p(W^2-M_p^2)}\int \sigma _T\sin \theta _{c.m.}\ d{\normalsize %
\theta _{c.m.}},
\end{equation}
}
\begin{figure}
\begin{center}
\plabel{fig5}
\let\picnaturalsize=N
\def\picsize{3in}
\def\picfilenamea{fig5.eps}
\ifx\nopictures Y\else{\ifx\epsfloaded Y\else\input epsf \fi
\let\epsfloaded=Y
\centerline{
\ifx\picnaturalsize N\epsfxsize \picsize\fi \epsfbox{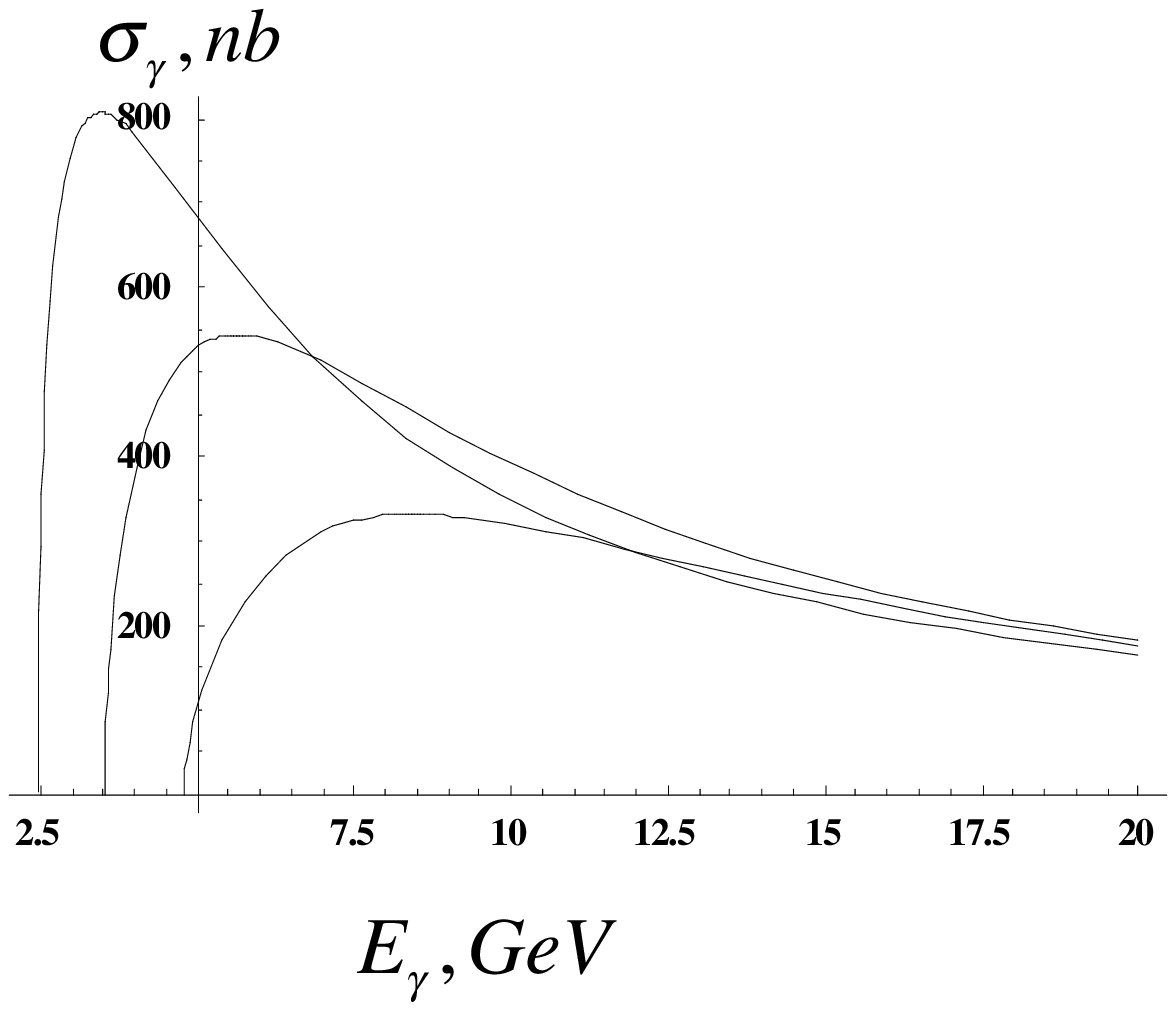}}}
\fi
\caption{Photoproduction cross--section in $nb$ as a function of
photon energy (in GeV) for various $\rhohat$ masses for standard parameters. From top to bottom
on the vertical axis this corresponds to $\rhohat$ masses of 1.4, 1.8 and
2.2 GeV.}
\end{center}
\end{figure}

The photoproduction cross--section $\sigma_{\gamma}$ is shown in
Figure 5. The cross--section peaks not far from the $\rhohat$
production threshold. The shape of the cross--section as a function of
photon energy is very similar for the test form factor. 

The reason for the fall in the photoproduction cross--section with
increasing photon energy is firstly that, as the photon energy increases, the
smallest allowed $|t|$ (where the cross--sections peak) decreases, so
that $q_{\alpha}^{\gamma}\approx q_{\beta}^{\rhohat}$ and the
factor $\epsilon_{\mu\nu\alpha\beta}
\epsilon_{\mu}^{\gamma}\epsilon_{\nu}^{\rhohat\ast}
q_{\alpha}^{\gamma} q_{\beta}^{\rhohat}$ in the amplitude vanishes.
Secondly, the $\gamma_5$ coupling of the $\pi^+$ to the proton and neutron is
such that it flips the spin of the nucleon. As
$t\rightarrow 0$ the proton and neutron 4--momenta become identical
and the spin flip would become zero, so that the amplitude $\sim t$
(as can be seen explicitly in Eq. \ref{dim}). This means that with
increasing photon energy the spin flip of the nucleon suppresses the cross--section.

We check the total cross--section as a function of
the wave function parameters of the participating conventional mesons
for 6 GeV photons and $\rhohat$ of mass 1.8 GeV. 

\begin{tabbing}
\hspace{8cm} \= Total cross--section ($nb$)\\
Standard parameters \>  540  \\
$\brho=0.45$ GeV and $\bpi=0.75$ GeV \protect\cite{kokoski87} \>  250
\end{tabbing}  
Hence, within this model revisions in the $\beta$'s and $\gamma_0$ can make
the cross--sections $\sim 25\%$ of the values quoted for photoproduction
cross--sections in Figure 5. 

We have already suggested that $\rho^+$ electro-- and photoproduction can test the
ideas in this work. Unfortunately the
relevant data for $\rho^+$ has not yet been taken and only $\rho^+$
inclusive photoproduction data exist \cite{omega84}. Photoproduction data is the most
likely to be forthcoming, and we show the dominant structure function
$\sigma_T$ in Figure 6. It may be observed that the structure function
is somewhat different from the $\rhohat$ structure function in Figure
2. This is mainly due to the fact that the mass of the $\rho$ is very
different from the $\rhohat$.  
We find that the $\rho$ structure functions
$\sigma_L,\; \sigma_{LT}$ and $\sigma_{TT}$ have similar parameter
dependence to their $\rhohat$ analogues.

\begin{figure}[t]
\plabel{fig6}
\let\picnaturalsize=N
\def\picsize{3in}
\def\picfilenamea{cap6.ps}
\ifx\nopictures Y\else{\ifx\epsfloaded Y\else\input epsf \fi
\let\epsfloaded=Y
\centerline{
\ifx\picnaturalsize N\epsfxsize \picsize\fi \epsfbox{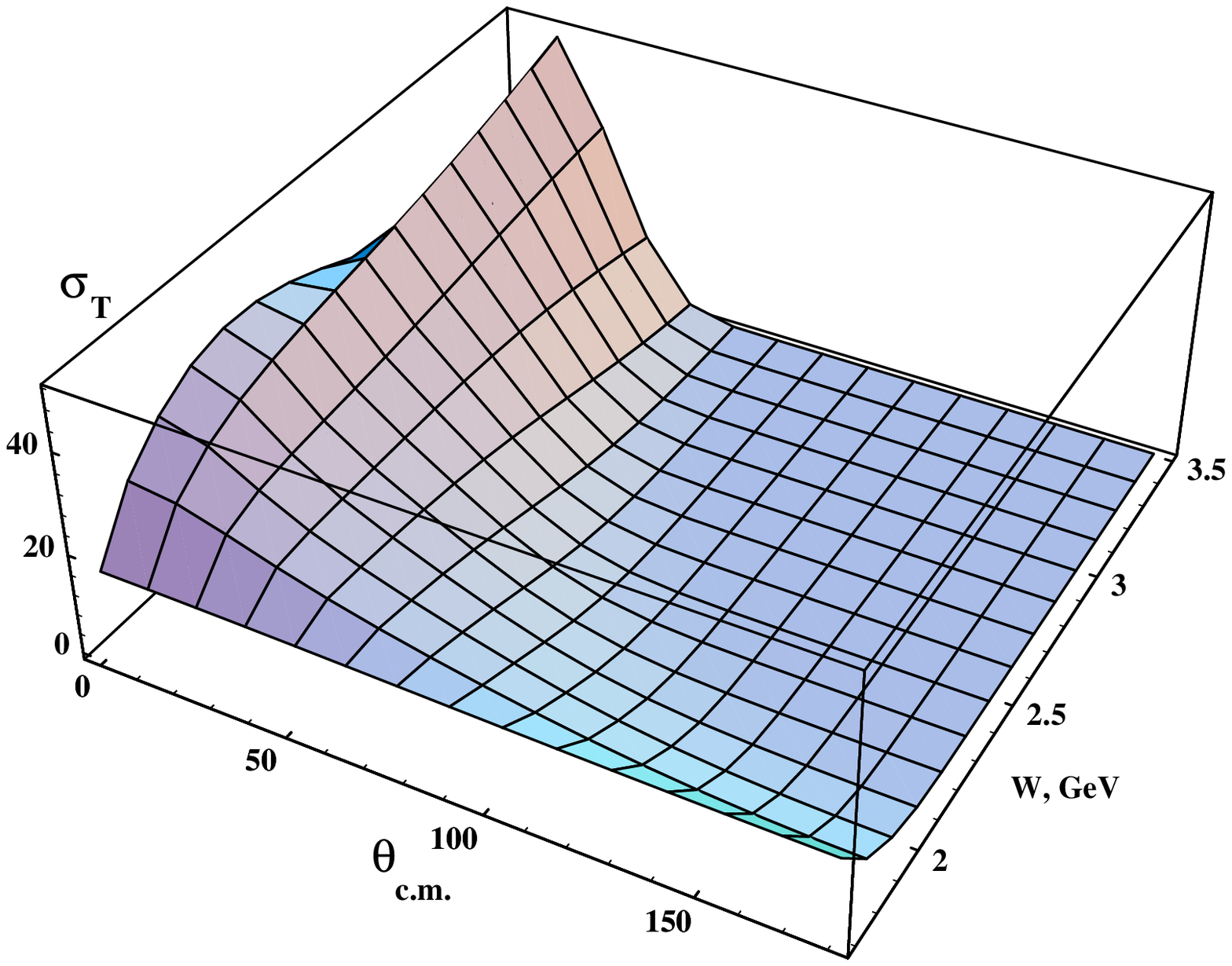}}}
\fi
\caption{The $W$ and $\theta_{c.m.}$ dependence of the structure
function $\sigma_T$ for $\rho^+$ production at $Q^2=0$. We vary $W$
within its kinematically allowed range for $E=6$ GeV. Within the framework of VDM, $\gamma^{\ast}$ couples
to an $\omega$, and the $\omega$ couples to the
$\rho^+$ and $\pi^+$ via G-parity allowed OZI allowed couplings. The structure functions are those of Eq.
\ref{dim} with all references to $\rhohat$ replaced by $\rho$. The
flux--tube model form factor for $\rho^+$ production is proportional to the form factor 
in Eq. \protect\ref{flux} with the understanding that all reference to $\rho$ is
replaced by $\omega$, and all reference to $\rhohat$ is
replaced with $\rho$. We use Eq. \protect\ref{flux} with $\beta_{\rho}=\beta_{\omega}=0.31$ GeV and $\beta_{\pi}=0.54$
GeV. The normalization of the structure function has hence been chosen to coincide with its
analogue in Figure 2 to facilitate comparison; and has no physical significance.}
\end{figure}

\section{Summary}

\hspace{.5cm} $\bullet$ We found that the electroproduction cross--section peaks at small
$Q^2$, $\theta_{c.m.}$ and large $W$, with the consequence that it is strongly enhanced  for small--angle
electron scattering.

$\bullet$ The kinematical
dependence of cross--sections only weakly depends on the
model--dependent form factor of the $\gamma\pi\to 1^{-+}$ transition. 
The conclusions drawn can also be tested in $\rho^+$ electro-- and photoproduction. 

$\bullet$ A Rosenbluth--type separation of electroproduction cross section and $\Sigma$--asymmetry 
measurements in photoproduction are necessary to verify the $\rhohat$ production mechanism.

 $\bullet$ The $1^{-+}$ photoproduction cross--section peaks at energies near to the reaction threshold and
reaches values around 0.3 to 0.8 $\mu$b depending on model parameters and the assumed mass 
of the $\rhohat$ meson.

\section{Conclusion}
We conclude that electro- and photoproduction of  $1^{-+}$ exotic mesons from a proton target has 
high enough cross sections  to be observed in forthcoming Jefferson Lab experiments.  Optimal conditions
to study $\rhohat$ photoproduction would require a high intensity beam of real (or quasi--real) photons with
variable energies between 2.5 and 10 GeV, assuming that the (still unknown) $\rhohat$ mass is within the range
of 1.4 to 2.2 GeV.

\vspace{.9cm}

\noindent {\bf Acknowledgements}

\vspace{.2cm}

Helpful discussions with G. Adams, A. Donnachie and S. Stepanyan are acknowledged.
We specifically thank Nathan Isgur for encouragement. The work of A.A. was supported by the US Department of Energy under
contract DE--AC05--84ER40150. P.R.P. acknowledges a Lindemann Fellowship from the English Speaking Union.

\appendix

\section{Appendix: Relationships between kinematical variables}

$Q^2$ and $W$ are related to $E^{'}$ and $\theta_e$ by

\beqn
Q^2 = 2 E E^{'} (1-\cos\theta_e)\hspace{1cm} W^2 =  -2 E E^{'}
(1-\cos\theta_e) + 2 M_p (E-E^{'}) + M_p^2
\eeqn
where $0\leq Q^2 \leq Q^2_{max}$ and $M_n+M_{\rhohat} \leq W \leq
\sqrt{M_p(M_p+2 E)}$; and $0 \leq \theta_e \leq \pi$ and $0 \leq E^{'}
\leq E_{max}^{'}$, with

\beqn\plabel{max}
Q^2_{max}  = \frac{2 E}{M_p+2E}(M_p^2+2 M_p E - W^2) \nonumber
\eeqn

\beqn
E_{max}^{'} = \frac{M_p E + \frac{1}{2} M_p^2 -
\frac{1}{2}(M_n+M_{\rhohat})^2}{E(1-\cos\theta_e) + M_p}
\eeqn

The variables $q_0^{\rhohat}, |{\bf q}^{\rhohat}|$ and $q_0^{\gamma},
|{\bf q}^{\gamma}|$ are defined in terms of $W$ and $Q^2$ by

\beqn
q_0^{\rhohat} = \sqrt{|{\bf q}^{\rhohat}|^2+ M_{\rhohat}^2} = \frac{W^2+M_{\rhohat}^2-M_n^2}{2W}\nonumber
\eeqn
\beqn
q_0^{\gamma} = \sqrt{|{\bf q}^{\gamma}|^2 -Q^2} = \frac{-Q^2+W^2-M_p^2}{2W}
\eeqn

${\bf p}_{\gamma}$ can be written in terms of $Q^2,\; W$ amd $t$ as

\beqn
{\bf p}_{\gamma}^2 =
\frac{M_{\rhohat}^4+Q^4+t^2+2M_{\rhohat}^2 Q^2-2M_{\rhohat}^2 t + 2Q^2 t}{4M_{\rhohat}^2}
\eeqn
where

\beqn\plabel{tdep}
t = (-q_0^{\gamma}+q_0^{\rhohat})^2-(|{\bf q}^{\gamma}|^2+|{\bf
q}^{\rhohat}|^2-2 |{\bf q}^{\gamma}| |{\bf
q}^{\rhohat}| \cos \theta_{c.m.})
\eeqn

For photoproduction, the photon energy is

\beqn
E_\gamma=\frac{W^2-M_p^2}{2 M_p}
\eeqn

\end{document}